\documentstyle[aas2pp4,psfig]{article}
\def\mcomp{\ifmmode M_{\rm C}\else$M_{\rm C}$\fi}
\def\rcomp{\ifmmode R_{\rm C}\else$R_{\rm C}$\fi}
\def\lcomp{\ifmmode L_{\rm C}\else$L_{\rm C}$\fi}
\def\mpsr{\ifmmode M_{\rm PSR}\else$M_{\rm PSR}$\fi}
\def\msun{\ifmmode M_\odot\else$M_\odot$\fi}
\def\rsun{\ifmmode R_\odot\else$R_\odot$\fi}
\def\teff{\ifmmode T_{\rm eff}\else$T_{\rm eff}$\fi}
\def\kms{\ifmmode {\rm km\,s^{-1}}\else$\rm km\,s^{-1}$\fi}
\def\psr{PSR~B1718$-$19}
\def\ngc{NGC~6342}

\def\Sref#1{\S\,\ref{sec:#1}}
\def\Fref#1{Fig.~\ref{fig:#1}}
\let\simgt\gtrsim
\let\simlt\lesssim
\let\internalcite\cite
\def\cite{\def\citename##1{##1}\internalcite}
\def\ctyr{\def\citename##1{}\internalcite}
\makeatletter
{\pt@width\wd\pt@box\box\pt@box\spew@ptblnotes%
\typeout{Page \the\pt@page\space of table \thetable\space has been set to
width \the\pt@width\space with \the\pt@nlines\space lines per page}%
\endcenter\end@float}
\def\startdata{\pt@line=0\pt@calcnlines%
\ifdim\pt@width>\z@\def\@halignto{to \pt@width}\else\def\@halignto{}\fi%
\let\fnum@table=\fnum@ptable\set@mkcaption%
\@float{table}\center\caption{\pt@caption}\leavevmode%
\setbox\pt@box=\pt@tabular{\pt@format}\pt@head}
\newenvironment{deluxetable*}[1]{\def\pt@format{\string#1}%
\set@tblnotetext\global\pt@ncol=0\global\pt@column=0\global\pt@page=1%
\def\pt@addcol{\global\advance\pt@ncol by\@ne}}%
{\pt@width\wd\pt@box\box\pt@box\spew@ptblnotes%
\typeout{Page \the\pt@page\space of table \thetable\space has been set to
width \the\pt@width\space with \the\pt@nlines\space lines per page}%
\endcenter\end@dblfloat}
\def\startdata{\pt@line=0\pt@calcnlines%
\ifdim\pt@width>\z@\def\@halignto{to \pt@width}\else\def\@halignto{}\fi%
\let\fnum@table=\fnum@ptable\set@mkcaption%
\@dblfloat{table}\center\caption{\pt@caption}\leavevmode%
\setbox\pt@box=\pt@tabular{\pt@format}\pt@head}
\def\thebibliography{\subsection*{REFERENCES}
\list{}{\labelwidth3em\leftmargin\labelwidth\labelsep\z@\parsep\z@
\itemsep\z@\itemindent-3em\usecounter{enumi}}
\def\refpar{\relax}
\def\newblock{\hskip .11em plus .33em minus .07em}
\sloppy\clubpenalty4000\widowpenalty4000
\sfcode`\.=1000\relax}
\makeatother
\psdraft
\begin{document}

\title{Optical Observations of the Binary
Pulsar System \psr:  Implications for Tidal Circularization}

\righthead{Optical Observations of \psr: Implications for Tidal
Circularization}

\author{M. H. van Kerkwijk\altaffilmark{1,2},
        V. M. Kaspi\altaffilmark{3}, 
        A. R. Klemola\altaffilmark{4},
	S. R. Kulkarni\altaffilmark{5},
        A. G. Lyne\altaffilmark{6},
	\and
        D. Van Buren\altaffilmark{7}}

\altaffiltext{1}{Astronomical Institute, Utrecht University, P. O. Box
                 80000, 3508 TA~~Utrecht, The Netherlands;
                 M.H.vanKerkwijk@astro.uu.nl}
\altaffiltext{2}{Institute of Astronomy, University of Cambridge,
                 Madingley Road, Cambridge, CB3~0HA, UK}
\altaffiltext{3}{Massachusetts Institute of Technology, Physics
		 Department, Center for Space Research 37-621, 70
		 Vassar Street, Cambridge, MA 02139;
		 vicky@space.mit.edu}
\altaffiltext{4}{UCO/Lick Observatory, University of California, Santa
                 Cruz, CA 95064; klemola@ucolick.org} 
\altaffiltext{5}{Palomar Observatory, California Institute of
                 Technology 105-24, Pasadena, CA 91125, USA; 
                 srk@astro.caltech.edu}
\altaffiltext{6}{Department of Physics, University of Manchester,
		 Jodrell Bank, Macclesfield, SK11 9DL, UK;
                 agl@jb.man.ac.uk}
\altaffiltext{7}{Infrared Processing and Analysis Center, California
		 Institute of Technology, Pasadena, CA 91125;
		 dave@ipac.caltech.edu}

\begin{abstract} 
We report on Keck and {\em{}Hubble Space Telescope} optical
observations of the eclipsing binary pulsar system \psr, in the
direction of the globular cluster \ngc.  These reveal a faint star
($m_{\rm{}F702W}=25.21\pm0.07$; Vega system) within the pulsar's
0\farcs5 radius positional error circle.  This may be the companion.
If it is a main-sequence star in the cluster, it has radius
$\rcomp\simeq0.3\,\rsun$, temperature $\teff\simeq3600\,$K, and mass
$\mcomp\simeq0.3\,\msun$.  In many formation models, however, the
pulsar (spun up by accretion or newly formed) and its companion are
initially in an eccentric orbit.  If so, for tidal circularization to
have produced the present-day highly circular orbit, a large stellar
radius is required, i.e., the star must be bloated.  Using constraints
on the radius and temperature from the Roche and Hayashi limits, we
infer from our observations that $\rcomp\simlt0.44\,\rsun$ and
$\teff\simgt3300\,$K.  Even for the largest radii, the required
efficiency of tidal dissipation is larger than expected for some
prescriptions.
\end{abstract}


\section{Introduction}\label{sec:intro}

\psr\ is a 1-s radio pulsar in the direction of the globular cluster
\ngc\ (\cite{lbhb93}).  It is in a 6.2\,h circular binary orbit with a
companion that has mass $\mcomp\geq0.11\,$\msun\ (assuming a pulsar
mass $\mpsr\simeq1.35\,\msun$).  At radio frequencies around
400--600\,MHz, irregular eclipses of the pulsar occur around superior
conjunction, indicating the presence of material around the companion.
The eclipsing material must be tenuous since the eclipses are absent
at higher frequencies.  The properties of the pulsar itself are not
special; from the spin period and its derivative, one infers a dipole
magnetic field strength
$B\simeq3.2\times10^{19}(P\dot{P})^{1/2}\simeq1.5\times10^{12}$\,G and
a spin-down age $\tau_{\rm{}sd}=P/2\dot{P}=10$\,Myr.

The system appears peculiar for several reasons (\cite{lbhb93}):
(i)~supernova activity in \ngc\ should have ceased long ago, hence one
does not expect it to contain an apparently recently formed pulsar;
(ii)~all other known binary pulsars in globular clusters have
millisecond periods and low magnetic fields, evidence that they were
``recycled'' or spun up by accretion from a binary companion, the
field presumably having decayed in the process (for a review, see
\cite{vdh95}); (iii) unlike in other eclipsing pulsars, the rotational
energy loss of \psr\ is many orders of magnitude below the critical
flux necessary to drive a wind from the companion, so the origin of
the eclipsing material is unclear.

Earlier papers on this system (\cite{lbhb93}; \cite{erg93};
\cite{wp93}; \cite{zwi93}; \cite{bk94}; \cite{esg96}) have focused on
the three peculiarities discussed above, mostly in the context of the
two favored formation scenarios: forming a new pulsar by
accretion-induced collapse of a white dwarf, and recycling an old
pulsar in a close encounter with other stars in the core of the
globular cluster.  We believe, however, another important issue is
that (iv) the current orbit is near-circular ($e\simlt0.005$), while
the formation should have left the pulsar and its companion in an
eccentric orbit.  An initially eccentric orbit is not only expected in
all formation models, but also indicated by the fact that the system
currently is not in the core of the cluster, as would be expected
given its mass, but offset by 2\farcm4; apparently, it received a
kick, which should have made the orbit eccentric as well.

It has been noted by Verbunt (\ctyr{ver94}) that if circularization
occurred, the energy dissipated in the companion would have been of
order the binding energy of the star.  Thus, the star might have
become bloated or even have been partially destroyed.  Little
attention, however, has been given to the question of whether the
tidal circularization efficiency is sufficient for circularization to
have happened within the spin-down age.

The above puzzles motivated us to try to identify the counterpart of
\psr.  In this article, we report on Jodrell Bank, Very Large Array
(VLA), Keck and {\em Hubble Space Telescope} ({\em HST\/}) observations
of \psr.  The improved position of the pulsar from the radio
observations and the results of the optical imaging observations are
reported and summarized in \Sref{obs}.  Given the crowded field,
establishing a precise tie between the radio and optical observations
is crucial.  A detailed account of our astrometry is given in
\Sref{astro}.  The photometry of a candidate object is also
reported in this section.  The constraints on the basic parameters of
the companion derived from optical and pulsar timing observations are
discussed in \Sref{constraints}.  In \Sref{disc}, we discuss
the implications of our results, focusing in particular on the
circularization.

\section{Observations}\label{sec:obs}

The position of the pulsar was first determined from radio timing
measurements using the 76-m Lovell telescope at Jodrell Bank
(\cite{lbhb93}).  We derived an improved timing position from data
obtained between 1994.6 and 1998.8:
$\rm\alpha_{J2000}=17^h21^m01\fs53$ and
$\rm\delta_{J2000}=-19\arcdeg36\arcmin36\arcsec$, with uncertainties
of $0\fs04$ and $6\arcsec$, respectively.  This position has a
relatively large uncertainty in declination because of the source's
proximity to the plane of the ecliptic.  An image of the field was
obtained on 11 November 1992 using the VLA in the A-configuration.  It
reveals a source within the timing error box with a flux density of
0.3\,mJy, consistent with the flux of the pulsar at the observing
frequency of around 1400\,MHz.  The position of this source is
$\rm\alpha_{J2000}=17^h21^m01\fs54$ and
$\rm\delta_{J2000}=-19\arcdeg36\arcmin36\farcs6$, with an uncertainty
of $0\farcs2$ in each coordinate.

\begin{figure}[t]
\epsfxsize0.9\columnwidth\epsfbox{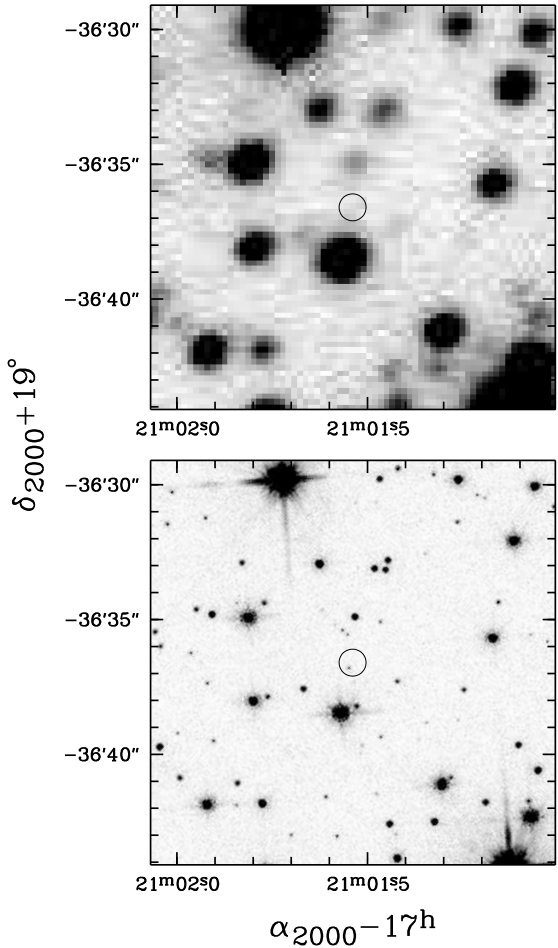}
\caption[]{Optical images of the field of \psr.  The top panel shows
part of the LRIS I-band image.  The exposure time was 240\,s and the
seeing 0\farcs9.  The limit to the flux of any star within the
0\farcs5 radius (95\% confidence) error circle centered on the VLA
position is $I>24.0$ (95\% confidence).  The bottom panel shows part
of the Planetary Camera image, taken through the the F702W filter.
The image shown is the median of all twelve 700\,s exposures.  The
candidate counterpart has
$m_{\rm{}F702W}=25.21\pm0.07$.\label{fig:images}}
\end{figure}

The \psr\ field was observed on 4 June 1995 using the Low-Resolution
Imaging Spectrometer (LRIS; \cite{occ+95}) on the 10-m Keck telescope.
Two series of exposures with integration times ranging from 10\,s to
240\,s were taken through several filters.  The seeing was
$\sim\!0\farcs9$ during the first series, and $\sim\!1\farcs2$ during
the second.  Our best-seeing I-band ($\sim\!0.8\mu{}m$) image is shown
in \Fref{images}.  These images showed that there was no
object in the VLA error box, and that with 95\% confidence
$I>24.0\,$mag.  Furthermore, they showed that a brighter star was so
close that it would be hard to make any further progress from the
ground.

The Wide Field Planetary Camera~2 (WFPC2) on board the {\em{}HST} was
used to observe the field through the F702W filter in four contiguous
{\em{}HST} orbits, from 14:39 to 20:07~{\sc{}ut} on 8 March 1997.  In
each spacecraft orbit, three 700-s exposures were taken, offset in
both X and Y by 0, 3, and 6 PC pixels in order to mitigate the effects
of hot pixels and imperfect flat fields.  The pointing was such that
the field around the pulsar is on a clean spot on the CCD of the
Planetary Camera~(PC).

The reduction and analysis started from the pipeline-calibrated PC
images (\cite{hhc+95}).  We first obtained median images for each of
the three observing positions.  The median image for the middle offset
was used for the astrometry described below.  Next, for each of the
three sets, we found pixels hit by cosmic rays in the individual
images by comparison with the median image; pixels with values more
than $5\sigma$ above the median were replaced with the median (here,
$\sigma$ is an estimate of the expected uncertainty based on the
median value).  Finally, the images were registered by applying
integer pixel shifts and co-added to form a grand average.  This
average is free of cosmic rays, but not of hot pixels and other chip
defects.  We used it for the photometry nonetheless, as our candidate
(see below) and the other stars we selected were on clean parts of the
chip.  For display purposes, however, we have used the median of all
registered images in \Fref{images}.

\section{Astrometry and Photometry}\label{sec:astro}

The astrometry of the \psr\ field was carried out in four stages.
First, 23 stars from the ACT Reference Catalog (\cite{ucw98}) were
used to derive an astrometric solution for a plate taken at epoch
1988.37 on Kodak 103a-G emulsion using the yellow lens of the 0.5-m
Carnegie double astrograph at Lick Observatory.  The model employed
five terms in each coordinate (proportional to 1, $x$, $y$, $xy$, and
$(x^2,y^2)$ in $(x,y)$).  The inferred rms error for a single star is
0\farcs16 in each coordinate, and the zero-point uncertainty in the
solution 0\farcs04.  We note that the errors are larger by about 50\%
than expected based on the measurement and ACT coordinate
uncertainties.  Probably, this is because some stars are blended with
fainter objects; thus, it should induce no systematic error.

Second, the plate solution was used to calculate right ascension and
declination for 30 fainter, relatively isolated stars in common with
the short LRIS R-band exposure.  For these stars, positions were also
measured on the LRIS frame, and corrected for instrumental distortion
using a bi-cubic transformation determined by J.~Cohen (1997, private
communication).  Solving for offset, scale and rotation, the inferred
rms single-star error is 0\farcs26 in each coordinate, consistent with
the expected measurement errors on the astrograph plate for these
relatively faint stars.  The zero-point uncertainty in the solution is
0\farcs05.

Third, using 23 fainter stars on a smaller part of the LRIS image
around the VLA position of \psr, the solution was transferred to one
of the 240\,s LRIS R-band images, solving only for the offset between
the two exposures.  The inferred rms error is $0\farcs011$ in each
coordinate, and the zero-point uncertainty in the solution 0\farcs002.

Finally, 37 stars were used to tie the astrometry to the WFPC2 PC
image (14 of these were used in the previous stage as well).  The
WFPC2 positions were corrected for instrumental distortion using the
cubic transformation given by Holtzman et al.\ (\ctyr{hhc+95}).
Offset, scale and rotation were left free in the solution\footnote{The
fitted scale is close to the value given by Holtzman et al.\
(\ctyr{hhc+95}), as is the offset between the fitted position angle
and the position angle of the {\em{}HST} V3 axis listed in the image
header.  If we fix the values, the inferred position of our proposed
counterpart changes by 0\farcs006 in RA and $-0\farcs005$ in Dec.
Thus, a perhaps more conservative estimate of the zero-point
uncertainty is about 0\farcs01 in each coordinate.}.  The inferred rms
error was 0\farcs017 in each coordinate, and the zero-point
uncertainty in the solution 0\farcs003.

From the numbers above, the total uncertainty in the zero points of
positions due to measurement errors is 0\farcs06 in each coordinate.
We expect systematic effects to be smaller than this; e.g., no large
systematic offset as a function of magnitude is seen on yellow
astrograph plates used for other applications.  A possible additional
uncertainty is the extent to which our ACT-based solution is on the
same system as the VLA position of \psr, i.e., the International
Celestial Reference System (ICRS).  The ACT combines the Tycho
(\cite{esa97}) and AC2000 (\cite{ucw+98}) catalogs to determine
positions and accurate proper motions.  The Tycho positions are on the
ICRS to within 0.6\,mas (\cite{esa97}), and any offset from the ICRS
at the plate epoch, which is close to the Tycho epoch (1991.25),
should be small as well.

In summary, we expect the uncertainties in the astrometry to be
dominated by measurement errors, 0\farcs06 in each coordinate for the
optical position, and 0\farcs2 for the radio position.  Combining the
two in quadrature, the 95\% confidence error radius is
$[-2\log(1-0.95)(0\farcs2^2+0\farcs06^2)]^{1/2}=0\farcs5$.  Within
this radius, there is one faint object (\Fref{images}), at
$\rm\alpha_{J2000}=17^h21^m01\fs549\pm0\fs004$ and
$\rm\delta_{J2000}=-19\arcdeg36\arcmin36\farcs76\pm0\farcs06$.  This
may be the optical counterpart of \psr.

We measured magnitudes for the candidate counterpart of \psr\ and for
a number of other stars in the field following the prescriptions of
Holtzman et al.\ (\ctyr{hbc+95}) and Baggett et al.\ (\ctyr{bcgr97}).
We performed aperture photometry for a range of different radii, and
used some thirty brighter stars in the frame to determine aperture
corrections relative to the standard 0\farcs5 (11\,pix) radius
aperture.  For the candidate, the best signal-to-noise ratio is for
relatively small apertures, with radii between 1.5 and 3\,pix.  From
these, we infer a count rate for the 0\farcs5 radius aperture of
$0.070\pm0.005{\rm\,DN\,s^{-1}}$ (we find consistent values for the
larger apertures; here, $1{\rm\,DN}$ corresponds to about 7 electrons,
given the gain used for our observations).  This count rate
corresponds to a magnitude $m_{\rm{}F702W}=25.21\pm0.07$ in the Vega
system (using $m_{\rm{}F702W}=22.428$ for a count rate of
$1{\rm\,DN\,s^{-1}}$ in the PC, and applying a $0.10\,$mag aperture
correction from 0\farcs5 radius to ``nominal infinity'';
\cite{bcgr97}).  Magnitudes for the individual frames show a standard
deviation of 0.25\,mag around the average, similar to what is found
for other stars at this brightness level.  We do not find modulation
at the 6\fh2 orbital period of \psr\ (which is well covered by our
observations), although the limit on the modulation amplitude is not
very restrictive: $<\!0.3\,$mag at 95\% confidence.

At or above the brightness level of the candidate counterpart, there
is about one object per four square arcseconds in this field.  Thus,
there is a probability of about one in five of finding an object
within the 95\% confidence error radius by chance.  If it is a chance
coincidence, the real counterpart must be substantially fainter.  We
derive a 95\% confidence limit of $m_{\rm{}F702W}=27.2$ for any other
object in the error circle (using the observed noise in the sky near
the VLA position, of
$\sigma_{\rm{}sky}=5.6\times10^{-4}{\rm\,DN\,pix^{-1}\,s^{-1}}$, and
the fact that within the 0\farcs5 radius error circle there are about
400 resolution elements).

\section{Observational Constraints}\label{sec:constraints}

We now evaluate the constraints set on the system by the observations.
We first summarize the constraints on the age of the system and the
mass of the companion set by radio observations, and then discuss the
constraints on the companion radius and temperature from our
{\em{}HST} detection and Keck limit.  We will assume that we detected
the companion, and that the system is located in \ngc.

\subsection{System Age}\label{sec:age}

Clearly, something happened to the system recently (as compared to the
cluster age).  One indication is the short inferred spin-down age of
the pulsar, $\tau_{\rm{}sd}=P/2\dot{P}=10$\,Myr.  This is the time
required for a dipole rotating {\it in vacuo} to spin down to the
present-day spin period from an infinitely fast rotation rate.  It is
thus an upper limit to the true age, unless other mechanisms
influenced the spin period (e.g., transient spin-up by accretion or a
braking index very different from that predicted by dipole emission).

Another indication of a recent event is that the system is offset from
the cluster core.  Unperturbed, a system as massive as this should
have settled in the core long ago due to mass segregation (in
$\sim\!0.5\,$Gyr given its present position).  Indeed, the progenitor(s)
of this system, which must have been massive as well, should have
resided in the core.  It seems natural to argue that it was a single
event that brought the system to its present state and kicked it out
of the core.

We note that it would require fine-tuning for the kick to have
resulted in the system remaining in a cluster orbit as wide as is
indicated by the 2\farcm4 offset (the half-mass radius is 0\farcm9;
for a discussion, see, e.g., \cite{phi92}).  If instead the system is
unbound, the kick needs to have happened $\simlt\!1\,$Myr ago, which
would imply that the event left the pulsar with a spin period only
slightly shorter than the present one.  This would be consistent with
scenarios in which the pulsar was spun up by accretion: for
Eddington-limited accretion, the equilibrium spin period is close to
1\,s (\cite{lbhb93}).  We conclude that the system has been in its
present state for 10\,Myr at most.

\subsection{Companion Mass}\label{sec:compm}

Radio pulse timing of \psr\ has yielded the mass function,
$f(\mcomp,\mpsr)=\mcomp^3\sin^3i/(\mcomp+\mpsr)^2=0.000706\,\msun$,
where $i$ is the inclination of the binary orbit (\cite{lbhb93}).
From the mass function, assuming $\mpsr\simeq1.35\,\msun$
(\cite{tc99}), one infers $\mcomp\geq0.11\,\msun$; furthermore, there
is a 95\% {\em{}a priori} probability that $i>18\arcdeg$ and hence
that $\mcomp<0.43\,\msun$.

The radio eclipses allow a direct constraint on $i$, albeit in a
model-dependent way.  Burderi \& King (\ctyr{bk94}) and Thorsett
(\ctyr{tho95}) calculated the expected attenuation of the radio flux
at different frequencies for a simple constant-velocity, spherically
symmetric wind, and found that they could reproduce the observations
of Lyne et al.\ (\ctyr{lbhb93}) for $i\simgt30\arcdeg$.  This would
correspond to $\mcomp\simlt0.25\,\msun$.  Burderi \& King argue that
almost certainly $i>20\arcdeg$, implying $\mcomp<0.35\,\msun$.  In
summary, most likely $0.11\simlt\mcomp\simlt0.35\,\msun$.

\subsection{Companion Radius and Temperature}\label{sec:randt}

The interpretation of the apparent F702W magnitude depends on the
companion radius \rcomp, effective temperature \teff, and (to a lesser
extent) surface gravity $\log{}g$, as well as the distance and
reddening to \ngc.  To determine the constraint set by our
measurement, we used parameters for \ngc\ from the May 1997 edition of
the catalog of globular clusters (\cite{har96}): reddening
$E_{B-V}=0.44$, distance modulus $(m-M)_V=16.15$, and metallicity
relative to solar ${\rm[Fe/H]}=-0.65$.  The distance scale used in the
catalog is similar to the Hipparcos-based one (e.g., that of Chaboyer
et al.\ [\ctyr{cdkk98}] would give $(m-M)_V=16.22$).  The reddening
corresponds to $A_V=1.36$, $A_R=1.02$, $A_I=0.65$ (using the
extinction curve of \cite{mat90}), and one infers $(m-M)_0=14.79$ and
$d=9.1\,$kpc.  We also used the evolutionary tracks for
$\rm[Fe/H]=-0.5$ stars of Baraffe et al.\ (\ctyr{bcah98}), to relate
temperatures and radii to absolute magnitudes in various standard
bands.

With these data in hand, we proceeded as follows.  First, we looked up
temperatures and corresponding radii for the main sequence (at age
10\,Gyr, appropriate for a metal-rich globular cluster; \cite{sw98}).
Second, we used $M_V$ and $M_R$ for these stars to calculate
$V_{\rm{}ms}$ and $R_{\rm{}ms}$ for the distance and reddening of
\ngc.  Third, for each $(V-R)_{\rm{}ms}$, we used the calibration of
Holtzman et al.\ (\ctyr{hbc+95}; Eq.~9, Table~10) to infer the
expected $R_{\rm{}F702W}$ magnitude corresponding to the observed
F702W count rate.  Fourth, we derived the radii required to match the
observations from the difference $R_{\rm{}F702W}-R_{\rm{}ms}$.  To
estimate the uncertainty, we assumed a total uncertainty of 0.3\,mag
in the magnitude difference, which we regard as a 95\% confidence
estimate (it is dominated by the estimated uncertainties in distance
and reddening).  Strictly speaking, one should redden the F702W
magnitude and then calculate $R_{\rm{}F702W}$ using $(V-R)_0$.
Furthermore, one should consider the influence of $\log{}g$ on the
star's colors, and take into account the slight difference in
metallicity.  None of these corrections, however, is important at the
present level of accuracy.

Another constraint is set by our I-band non-detection.  This
corresponds to an upper limit to the $m_{\rm{}F702W}-I$ color, and
thus a lower limit to the effective temperature.  We find
$\teff>3150\,$K for $m_{\rm{}F702W}-I>1.4$ (here, we have decreased
the limit to the magnitude difference by 0.2\,mag to account for
uncertainties in $m_{\rm{}F702W}$, in the color transformation, in the
effects of changes in $\log{}g$, and in the reddening).

The observational constraints can be summarized graphically in a
diagram of \rcomp\ versus the effective temperature \teff\ of the
companion (see \Fref{teffvsr}).  To make further progress we need to
have some knowledge of the companion's nature.  We consider two cases:
the companion is an ordinary main sequence star; and the companion is
a bloated star.

\begin{figure}[t]  %***********  FIGURE 2 ************
\epsfxsize0.95\columnwidth\epsfbox{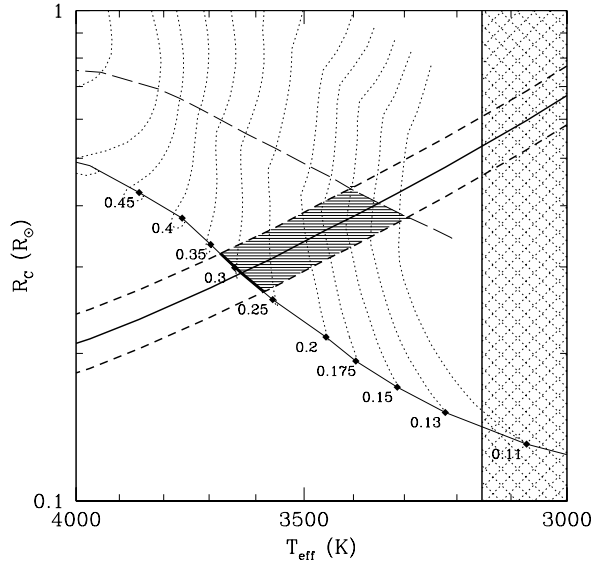}
\caption[]{Constraints on the radius and temperature of the companion
of \psr.  The continuous curve with dashed curves next to it indicates
the constraint set by our measured F702W magnitude.  The bold vertical
line with the hashed region to the right indicates the limit to the
temperature set by the limit to the $m_{\rm{}F702W}-I$ color.  The
continuous curve labeled with masses (in solar units) indicates the
relation expected for main sequence stars with $\rm[Fe/H]=-0.5$; the
bold section of this curve represents the allowed range if the star is
on the main sequence (\S\ref{sec:mscomp}).  As argued in the text, the
companion may have become bloated due to irradiation or tidal
dissipation.  The dotted lines indicate pre-main-sequence tracks for
various masses.  These are close to the Hayashi limit, and stars in
hydrostatic equilibrium can only be on or to the left of these tracks.
The long-dashed curve connects the maximum (Roche) radii the companion
could have if it were on such a track.  The shaded region shows the
full range of allowed parameters (\Sref{bloat}).\label{fig:teffvsr}}
\end{figure}       %***********  FIGURE 2 ************

\subsubsection{A Main Sequence Companion}\label{sec:mscomp}

In \Fref{teffvsr}, the locus of radius versus effective temperature
for main sequence stars of different masses is indicated (from
\cite{bcah98}).  From the intersection with the region allowed by our
observation, one infers that if the companion were a main sequence
star, it would need to have $0.26\simlt\rcomp\simlt0.33\,\rsun$ and
$3580\simlt\teff\simlt3680\,$K.  Its mass would be
$0.26\simlt\mcomp\simlt0.34\,\msun$, consistent with the constraints
inferred dynamically and from eclipse modeling (\S\ref{sec:compm}).

\subsubsection{A Bloated Companion}\label{sec:bloat}

The companion does not necessarily have to be on the main sequence.
Indeed, bloating might be expected: most formation mechanisms produce
a binary that has substantial eccentricity initially, and the amount
of energy that needs to be dissipated in order to circularize the
orbit is a substantial fraction of the binding energy of a low-mass
star\footnote{The star cannot be large because of having evolved off
the main-sequence, as a star at or past the turn-off mass would have a
luminosity much higher than the limit implied by our observations.
Even if mass were lost, the thermal timescale for such a star likely
is too long for the luminosity to have changed substantially.}
(\cite{ver94}).  Bloating may also be required, as alluded to in
\Sref{intro} and discussed in \Sref{disc}, for tidal
dissipation to have circularized the orbit in the short time since
formation.

We are not aware of detailed calculations of bloating due to tidal
dissipation for almost completely convective, low-mass
stars\footnote{See Podsiadlowski (\ctyr{pod96}) for calculations for
somewhat more massive stars, for which only the outer layers are
convective.}.  Regardless of the expansion process, however, for a
given mass an upper limit to the radius is set by the Roche radius,
above which mass transfer would occur.  Furthermore, for given mass
and radius, a lower limit to the temperature is set by the Hayashi
(\ctyr{hay61}) limit, below which a star cannot be in hydrostatic
equilibrium.

To delineate the constraint on the temperature, we can use
pre-main-sequence tracks, since these follow the Hayashi limit
closely.  Indeed, it seems not unlikely that the companion is
currently contracting along a similar track, since all processes that
could have induced bloating (irradiation, tidal heating) should have
ceased to be operative (the pulsar's spin-down luminosity being very
low and tidal dissipation having ceased as the orbit became circular).
Of course, it is not clear that there has been enough time for the
companion to adapt to a quasi--pre main sequence configuration.  On
the other hand, if the bloating is due to tidal heating, and if the
energy was dissipated somewhere in the convective regions, convection
would have distributed the energy throughout the star and expansion
may well have been along a quasi--pre main sequence track as well.

In \Fref{teffvsr}, dotted lines indicate the pre-main-sequence tracks
of Baraffe et al.\ (\ctyr{bcah98}), and the long-dashed line shows the
Roche radii for the whole range of masses.  Under our assumptions, the
companion must lie between the latter line and the main sequence, and
at any given radius and temperature, an upper limit to its mass is set
by the mass for which the pre-main-sequence track passes through that
radius and temperatures.  In addition, it must reproduce the observed
F702W flux and be consistent with the $m_{\rm{}F702W}-I$ limit.  The
allowed ranges are $3300\simlt\teff\simlt3680\,$K and
$0.44\simgt\rcomp\simgt0.26\,\rsun$.  The inferred masses are
$0.11\simlt\mcomp\simlt0.34\,\msun$, consistent with the constraints
inferred dynamically and from eclipse modeling (\S\ref{sec:compm})
except for the very low-mass end ($\mcomp\simlt0.12\,\msun$), at which
total eclipses would be expected (for a 1.35\,\msun\ neutron star),
which are not observed.

\section{Discussion}\label{sec:disc}

We now discuss the implications of our results.  The assumption we
continue to make is that the system is associated with \ngc\ and that
we have detected the counterpart.  The constraints we regard as most
important are: (i) the system was brought to its present state
$\simlt\!10\,$Myr ago; (ii) the current orbit is nearly circular
($e\simlt0.005$); and (iii) the companion has
$0.26\simlt\rcomp\simlt0.44\,\rsun$ and
$3680\simgt\teff\simgt3300\,$K.  In these ranges, the left-hand limits
correspond to the case where the companion is a $\sim\!0.3\,\msun$
main-sequence star, while the right-hand limits correspond to the case
where the companion is a $\sim\!0.13\,\msun$ star which has been
maximally bloated and is currently contracting along a
pre-main-sequence track.

Below, we first briefly review the formation models that have been
suggested, in order to set the stage for a discussion of how the
system could have been circularized.  We end by noting briefly what
constraints one could set if the star we detected is not the
counterpart of \psr, or if the system is not in \ngc.

\subsection{Formation Models}\label{sec:form}

Two models have been suggested to explain the origin of the \psr\
system.  One is that the neutron star was formed early in the life of
the globular cluster (\cite{lbhb93}; \cite{wp93}; \cite{zwi93}; Ergma
et al.\ \ctyr{esg96}).  It had stopped being a radio pulsar and was
dormant until it had a close interaction with a star or binary,
$\sim\!10\,$Myr ago.  During that interaction, some mass was lost from
a normal star, part of which was accreted by the neutron star.  As a
consequence, it was spun up sufficiently for the radio mechanism to
become active again, yet did not undergo a long phase of mass transfer
and thus kept a large magnetic field.  Similar close encounters have
been invoked to explain the presence of two other long-period (but
single) pulsars in globular clusters: PSR~B1745$-$20 in NGC~6440 and
PSR~B1820$-$30B in NGC~6624 (\cite{lmd96}).  

To give the system the velocity required by the present location far
outside the cluster core and conserve momentum, the putative
interaction must have involved at least one other object.  The
system's initial orbit is expected to have been highly eccentric,
$e\simgt0.8$ (see, e.g, \cite{phi92}).

The second model that has been considered for the origin of the \psr\
system is that the neutron star was formed via accretion-induced
collapse (AIC) of a white dwarf (\cite{lbhb93}; \cite{erg93};
\cite{wp93}; Ergma et al.\ \ctyr{esg96}).  The kick imparted to the
neutron star during AIC, whether intrinsic or due to mass loss, must
have been relatively small, since otherwise the systemic velocity
would have been so large that the system would have left the cluster
long ago.  For a kick of order the $20{\rm\,km\,s^{-1}}$ escape
velocity from the core (\cite{web85}), an initial eccentricity of
$\sim\!0.2$ is expected.

Given such a kick, the increase in the orbital separation should have
been small, about a factor 1.25.  Thus, the companion, which must have
filled its Roche lobe before AIC for mass transfer to occur, should
still be close to filling its Roche lobe.  This implies that it cannot
have been a Roche-lobe filling main-sequence star, as this would
require a mass of $\sim\!0.7\,\msun$ and hence a much brighter optical
counterpart than we observe (see \Fref{teffvsr}).  It has been
suggested that the star was bloated already before AIC due to
irradiation from the primary during the mass transfer phase
(\cite{erg93}; Ergma et al.\ \ctyr{esg96}).  It is not clear whether
strong bloating is possible by irradiating only one side of a star
(e.g., \cite{kfkr96}), but if it happened, the star should still be
bloated, because the thermal time scale is much longer than the pulsar
characteristic age.

\subsection{Circularization}\label{sec:circ}

In both formation scenarios that have been suggested, the initial
binary orbit is expected to be eccentric.  Indeed, in general any
event that imparted a systemic velocity large enough for the system to
(almost) escape the cluster will most likely have left the binary
orbit eccentric.  The current tiny eccentricity therefore requires
explanation.  Since eccentricity likely decays exponentially, the
current low $e$, even for an initial eccentricity as small as
$e\simeq0.1$, requires a circularization time
$t_{\rm{}circ}\simlt\tau_{\rm sd}/4=2.5\,$Myr.

It is not straightforward to estimate whether such a short $t_{\rm
circ}$ is possible.  This is because the companion is probably
completely convective, with convective turnover time scales far longer
than the 6.2-h orbital period (see below).  This makes the energy
transfer less efficient, and thus circularization time scales longer,
but it is not clear to what extent.  We will first use the
prescription of Zahn (\ctyr{zah89}), in which the efficiency is
assumed to decrease linearly with the ratio of the convective to
orbital timescale, and then discuss the prescription of Goldreich \&
Nicholson (\ctyr{gn77}) and Goodman \& Oh (\ctyr{go97}), in which the
efficiency decreases almost quadratically with the timescale ratio.
We should stress that at present it is not clear that either
prescription is reliable; see Goodman \& Oh for a discussion.

Following the formalism of Zahn (\ctyr{zah89}; Eq.~[21]), we
write\footnote{See Phinney (\ctyr{phi92}) for a pedestrian
derivation.} 
\begin{equation}
\frac{1}{t_{\rm circ}}=-\frac{1}{e}\frac{{\rm d}e}{{\rm d}t} =
21\frac{\lambda_{\rm circ}}{t_{\rm f}}q(1+q)\left(\frac{\rcomp}{a}\right)^8,
\end{equation}
where $t_{\rm{}circ}$ is the circularization timescale, $e$ the
eccentricity, $\lambda_{\rm{}circ}$ a dimensionless average of the
turbulent viscosity weighted by the square of the tidal shear,
$t_{\rm{}f}=(\mcomp\rcomp^2/\lcomp)^{1/3}$ the convective friction
time, $q=\mpsr/\mcomp$ the mass ratio, and $a$ the orbital separation.
Here, all uncertainty is hidden in the parameter
$\lambda_{\rm{}circ}$.  In the prescription of Zahn (\ctyr{zah89}), it
can be approximated by
$\lambda_{\rm{}circ}\simeq0.019\alpha^{4/3}(1+\eta^2/320)^{-1/2}$,
where $\alpha$ is the mixing length parameter and
$\eta=2t_{\rm{}f}/P_{\rm{}orb}$ a measure of the timescale mismatch.

For a main-sequence star companion with parameters in the ranges
listed above, we find $t_{\rm{}f}\simeq0.5\,$yr and $\eta\simeq1500$.
Assuming $\mpsr=1.4\,\msun$ (i.e., $a\simeq2\,\rsun$) and taking
$\alpha=2$ (as in Zahn [\ctyr{zah89}]; consistent with the
observational constraint derived by Verbunt \& Phinney [\ctyr{vp95}]
for small $\eta$), one infers
$\lambda_{\rm{}circ}\simeq6\times10^{-4}$, and a circularization time
in the range $5\simlt{}t_{\rm{}circ}\simlt17\,$Myr, too long to
understand the current small eccentricity.  If the star is bloated,
circularization is much faster, because of the very strong dependence
on the ratio $\rcomp/a$: we find $t_{\rm{}circ}\simeq0.07\,$Myr for a
maximally bloated companion.

The situation is different for the prescription of Goodman \& Oh
(\ctyr{go97}), in which $\lambda_{\rm{}circ}\propto\eta^{-2}$.
Extrapolating in their Fig.~2, we infer
$\lambda_{\rm{}circ}\simeq4\times10^{-6}$ (note their slightly
different definition of $\eta$).  Thus, the inferred circularization
times are two orders of magnitude longer than those estimated with the
prescription of Zahn (\ctyr{zah89}); if correct, it may be difficult
to understand how the orbit could have been circularized even if the
companion was maximally bloated.

As mentioned in \Sref{constraints}, the star could have become
bloated due to the energy dissipated by the circularization proper
(\cite{ver94}).  It is difficult to estimate, however, by how much, as
it is not clear where, how, and on what time scale the tidal energy is
dissipated, especially in such a low-mass star.  For high
eccentricity, the tides excite low-order pulsation modes in the star,
which will be dissipated, either by direct viscous damping
(\cite{koc92}), or, perhaps more likely, by non-linear coupling to
higher-degree modes and damping of these (\cite{kg96}).  For either
case, it appears that for a low-mass, (almost) completely convective
star ($\simlt\!0.5\msun$), most of the energy will be dumped in the
outermost layers.  These will be heated and may expand, which would
lead to stronger tidal coupling.

If the expansion involved the whole star, or a substantial fraction of
it, most likely the star would still be bloated, as the thermal time
would be much longer than the pulsar characteristic age.  Also if only
the outer layers of the star expanded, however, it seems likely the
star would still be bloated, as otherwise it would be difficult to
reduce the eccentricity sufficiently.  This is because the tidal
luminosity will decrease rapidly with decreasing eccentricity and
increasing periastron distance; if the expanded layers shrunk too
quickly in response, the circularization time would have become long
again while the eccentricity was still substantial.  One way to verify
whether the companion is still bloated is to measure its temperature.

\subsection{Caveat: Fainter Companion}\label{sec:caveat}

There is a probability of about one in five of a chance coincidence
between \psr\ and the object in our HST images (\Sref{astro}).  If
so, the companion has to be substantially fainter, with
$m_{\rm{}F702W}>27.2$ (\Sref{astro}).  One can go through a
similar exercise as in \Sref{randt} to constrain the companion
properties for this case.  The result is that the companion needs to
have mass $\simlt\!0.15\,\msun$, close to the minimum allowed by
pulse timing.  In this case, the circularization time scale problem
is exacerbated.

Another possibility is that \psr\ is not associated with \ngc.
Indeed, the dispersion measure is only $71{\rm\,cm^{-3}\,pc}$, while
$130{\rm\,cm^{-3}\,pc}$ is expected for \ngc\ (\cite{tc93}).  Taken at
face value, a distance of only 3\,kpc is implied.  If \psr\ were in
the foreground, again the companion would be less luminous and thus
less massive, even more problematic for circularization.

\section{Conclusions}\label{sec:conclusions}

In summary, using {\rm{}HST} observations, we have detected a faint
star at the position of the unusual eclipsing binary radio pulsar
\psr.  This faint star most likely is the pulsar's companion.  We have
shown that it is difficult to explain the highly circular present-day
orbit if the companion is a $\mcomp\simeq0.3\,\msun$,
$\rcomp\simeq0.3\,\rsun$ main-sequence star (unless circularization
was not by tidal interaction).  If it is a bloated,
$\rcomp\simgt0.4\,\rsun$ star, circularization may be sufficiently
rapid, depending on the extent to which the efficiency of tidal
dissipation is suppressed by the orbital period being far shorter than
the convective turnover time in the star.  The system thus provides
an interesting test-case for tidal-interaction theory.

% Three possible formation mechanisms for the system are: (i) a resonant
% interaction in the core involving at least three stars, with the
% encounter or tidal interaction leaving the companion bloated so that
% the orbit could be circularized; (ii) a resonant encounter involving
% four stars, in which one star was destroyed, and the drag from its
% remainders helped to circularize the orbit; (iii) AIC of a massive
% white dwarf accreting from a low-mass, strongly bloated companion.

A measurement of the color of the companion would be the best way to
determine whether it is bloated or not.  For a main-sequence star with
$\teff=3640\,$K and a bloated star with $\teff=3300\,$K, one expects
$(R-I,I-J,J-H,H-K)_0=(1.0,1.1,0.6,0.2)$ and $(1.3,1.4,0.6,0.2)$,
respectively.  Radio and optical proper-motion studies could settle
association of our candidate counterpart with \psr, and of the system
with \ngc.

\acknowledgements We are indebted to the referee, R. Webbink, for his
thoughtful report, and, in particular, for pointing out the importance
of the Hayashi limit for constraining the temperature of a bloated
companion.  We also thank him for the estimate of the relaxation time
of \psr\ at its present position in \ngc.  We thank M.\ Goss for help
with the VLA observations, and P. Goldreich, V. Kalogera, F. Rasio,
S. Sigurdsson, C. Tout, M. van den Berg, F. Verbunt, and R. Wijers for
useful discussions.  The observations were obtained at the W.~M.~Keck
Observatory on Mauna Kea, Hawaii, which is operated by the California
Association for Research in Astronomy, and with the NASA/ESA Hubble
Space Telescope at STScI, which is operated by AURA.  The reduction of
the optical data was done using the Munich Image Data Analysis System
(MIDAS), which is developed and maintained by the European Southern
Observatory.  This research made use of the SIMBAD data base.  We
acknowledge support of a NASA Guest Observer grant (GO-06769.01-95A),
a fellowship of the Royal Netherlands Academy of Arts and Sciences
(MHvK), a NSF grant (95-30632; ARK), a Sloan Research Fellowship
(VMK), and visitor grants of the Netherlands Organisation for
Scientific Research NWO (VMK) and the Leids Kerkhoven-Bosscha Fonds
(MHvK, VMK).  MHvK thanks MIT and Caltech for hospitality, VMK the
Aspen Center for Physics and Utrecht University.

\end{document}